\documentclass[preprint2]{emulateapj}
\def\simlt{\lower.5ex\hbox{$\; \buildrel < \over \sim \;$}}
\def\simgt{\lower.5ex\hbox{$\; \buildrel > \over \sim \;$}}

\def\cm3{{\rm cm^{-3}}}
\def\kms{km s$^{-1}$}
\def\msol{{$M_\odot$}}
\def\lsol{{$L_\odot$}}

\def\alphaJ{{\alpha_{2000}}}
\def\deltaJ{{\delta_{2000}}}

\def\g54{G54.1+0.3}
\def\akari{{\em AKARI}\ }
\def\spitzer{{\em Spitzer}}

\def\um{$\mu$m}

\def\fe164{[Fe II] 1.644~$\mu$m}

\def\h212{H$_2$ 2.122~$\mu$m}
\def\schi{{\sc Hi}\ }

\def\tmass{{2MASS}}
\def\jhk{{$JHK_s$}}
\def\jmh{{$J-H$}}
\def\hmk{{$H-K_s$}}
\def\emt{[8]$-$[24]}
\def\tmf{[3.6]$-$[5.8]}
\def\teff{T_{\rm eff}}

\shortauthors{Koo et al.}
\shorttitle{Star-Forming Infrared Loop around G54.1+0.3}

\begin{document}
\title{A Massive-Star Forming Infrared Loop around 
the Crab-like Supernova Remnant G54.1+0.3:
Post Main-Sequence Triggered Star Formation?}

\author{Bon-Chul Koo\altaffilmark{1}, Christopher F. McKee\altaffilmark{2}, 
Jae-Joon Lee\altaffilmark{3}, Ho-Gyu Lee\altaffilmark{1}, Jeong-Eun Lee\altaffilmark{4}, 
Dae-Sik Moon\altaffilmark{5}, Seung Soo Hong\altaffilmark{1}, Hidehiro Kaneda\altaffilmark{6}, 
and Takashi Onaka\altaffilmark{7}}

\altaffiltext{1}{Department of Physics and Astronomy, Seoul National University, Seoul 151-742, Korea; 
koo@astrohi.snu.ac.kr, hglee@astro.snu.ac.kr, sshong@astro.snu.ac.kr}
\altaffiltext{2}{Department of Physics and Astronomy, University of California, Berkeley, CA94720, USA;
cmckee@astro.berkeley.edu}
\altaffiltext{3}{Astronomy \& Astrophysics Department, Pennsylvania State University, University Park PA16802, USA; lee@psu.edu}
\altaffiltext{4}{Department of Astronomy and Space Science, Sejong University, Seoul 143-747, Korea;
jelee@sejong.ac.kr}
\altaffiltext{5}{Department of Astronomy \& Astrophysics, University of Toronto, Toronto ON M5S 3H4, Canada;
moon@astro.utoronto.ca}
\altaffiltext{6}{Institute of Space and Astronautical Science, JAXA,
Kanagawa 229-8510, Japan; kaneda@ir.isas.jaxa.jp}
\altaffiltext{7}{Department of Astronomy, 
University of Tokyo, Bunkyo-ku, Tokyo 113-0003, Japan; onaka@astron.s.u-tokyo.ac.jp}

\begin{abstract}

We report the discovery of a star-forming loop around the
young, Crab-like supernova remnant (SNR) \g54\ using the \akari\ infrared satellite.
The loop consists of 
at least
11 young stellar objects (YSOs) embedded in a
ring-like diffuse emission of radius $\sim 1'$. 
The YSOs are bright in the mid-infrared and 
are also 
visible in the \spitzer\ Space Telescope Galactic plane survey images.  
Their \spitzer\ colors are similar to those of Class II YSOs in 
\tmf\ but significantly redder in \emt, i.e., 
$0<$\tmf$<1.2$ and $5<$\emt$<9$.
Most of them have near-infrared counterparts in the \tmass\ \jhk\ images, and 
some of them have an optical counterpart too.
Their \jhk\ colors and magnitudes indicate that the YSOs are 
massive ($\simgt 10$~\msol) pre-main-sequence stars 
at the same distance to the SNR, i.e., 8 kpc, which 
supports the association of the star-forming loop with the SNR.
The dereddened spectral energy distributions are 
similar to 
{\em early} 
Herbig Be stars, which are early B type pre-main-sequence 
stars with inner disks that have been destroyed.
The confinement to a loop structure indicates that the YSOs
are young, i.e., 
$\simlt 2$ Myr.
We propose that their formation is triggered by 
the progenitor star of \g54,
which has a mass of $\simlt 15$ \msol. 
The triggering must have occurred near the end of the progenitor's life, 
possibly
 after it had evolved off the main sequence.

\end{abstract}


\section{Introduction}

Massive stars can trigger the formation of a next generation of 
stars. Their strong stellar ultraviolet radiation and winds
can compress the ambient medium so that it becomes gravitationally
unstable and collapses.
Numerous examples supporting the scenario of
triggered star formation have been proposed, from
young stars formed in small bright-rimmed globules inside HII regions to
OB associations around
kiloparsec supershells
\citep[see][and references therein]{elm02a, zin07}.

In this Letter, we report the discovery of a
unique example of
triggered star formation, namely triggering by 
the progenitor star of a young, core-collapse SNR.
The SNR (\g54)
is Crab-like, with centrally-brightened
synchrotron emission in radio and X-rays \citep{vel88, lu02}.
The extent of the remnant is $120''\times 80''$ and is
much fainter than the Crab, e.g., 0.5 Jy versus 1040 Jy at 1 GHz.
At the center, a 136 ms radio/X-ray pulsar
with a characteristic age of 2900 yr has been
discovered \citep{lu02}.
The distance to the SNR is 8 kpc (\S~4).
This object provides a laboratory in which to
explore the physics of triggered star formation.

\section{Discovery of the IR Loop}

A mid-infrared (MIR) 
observation of \g54\ was done using the  
Infrared Camera (IRC) aboard \akari\ on 2007 April 17. 
The observation was performed in a framework of the 
\akari\ mission program: 
the ISM in our Galaxy and nearby galaxies 
(ID: 1401070.1, PI: H.K.)
The IRC is equipped with three wave band channels covering the 2.6--26 $\mu$m 
wavelength range 
with a 10$'$ $\times$ 10$'$ field-of-view (FOV) \citep{ona07}.
We used the MIR-L channel in IRC02 mode, which
gave two band images, L15 and L24, centered at wavelengths of
15.58 and 22.89 $\mu$m, 
respectively.
Their angular resolutions are $5.''7$ (L15) and $6.''8$ (L24).
The total on-source integration time was 196 s for each band.
The basic calibration and data handling
were processed using the
standard IRC Imaging Data Reduction Pipeline version 070104
\footnote{http://www.ir.isas.jaxa.jp/ASTRO-F/Observation/DataReduction/IRC/}.
The positional uncertainty is $0.''15$ at the 1-$\sigma$ confidence level.

Fig. 1 shows the \akari\ L15 image of \g54
with the 4.85 GHz radio continuum image overlaid.
The radio image was made from archive data of the
Very Large Array (VLA) of the National Radio Astronomy Observatory
\footnote{The National Radio
Astronomy Observatory is a facility of the National Science Foundation
operated under cooperative agreement by Associated Universities, Inc.};  the data were
originally obtained by \cite{vel88}.
The \akari image
shows a well-defined, bright, ring-like structure at the position of
the SNR \citep[see also][]{sla07}.
The ring appears partially complete with its northeastern portion
opened, with faint ridges extending out from
the ends of the bright ring.
The bright portion of the ring (hereafter the ``main ring")
is elongated along the northwest-southeast direction and has
an extent of $\sim 105''\times 54''$,
or $4.1\times 2.1$ pc at 8 kpc.
The ring is composed of both diffuse and compact sources.
Most of the compact sources are point-like except the
bright one at the northwestern end.
There are more than 10 point sources distributed along the main ring
and several others at the `tips' of the faint ridges.
Most of them are bright in the 24 \um\ image and have colors 
quite different from foreground/background stars.
This can be seen in Fig. 1 (right), which  
is a three-color image produced from the \spitzer\ 
IRAC 5.8 \um\ (B), \akari 15 \um\ (G), 
and \spitzer\ MIPS 24 \um\ (R) images. 
We obtained the \spitzer\ image of the source
from the GLIMPSE (Galactic Legacy Infrared Mid-Plane Survey Extraordinaire,
Release II) and MIPSGAL legacy projects. 
There are 11 sources with colors significantly redder than 
the other stars in the field (see also the inset in Fig. 2).

\section{Young Stellar Objects in the IR Loop}

We conducted photometry of a $4.'5$-radius area surrounding the
IR loop using the \akari\ 15 and 24 \um\ and the \spitzer\ MIPS 24 \um\ 
and 70 \um\ images. 
We have identified 151 sources detected in at least one of the 
15 or 24 \um\ bands; most
of them were detected at 15 \um\ while 76\% were detected at 24 \um.
At 70 \um, only one source was identifiable and 
only upper limits are derived for the rest.
The photometry of stars in the IR loop is not 
straightforward because of 
the bright diffuse emission from the loop. 
We did PSF photometry 
to minimize the contamination due to the diffuse emission.
Sources are crossidentified with the \spitzer\ GLIMPSE catalog, which contains
fluxes of the four IRAC bands,
and also with the 2MASS 
Second Incremental Release Point Source Catalog.

Fig. 2 shows the \spitzer\ \tmf\ versus \emt\ color-color plot 
of the 64 
sources detected in the IRAC and MIPS 24 \um\ bands 
with a [24] magnitude error of less than 0.2 mag. 
The stars clustered near (0.0, 0.0) are 
probably main-sequence or giant stars along the line of sight.
The sources along the IR loop are well defined in this diagram 
by their strong \emt\ excess and moderate or small \tmf\ excess, i.e.,
$5<$\emt$<9$ and $0<$\tmf$<1.2$. 
The sources within this box are marked by filled circles.
The half-filled circles with arrows in the box represent the 
sources not detected in one of the IRAC 5.8 and 8.0 \um\ bands but 
are included because they are likely to fall into the box based on their MIR colors.
We marked their positions using 
a shorter IRAC band, e.g., if a source is not detected 
at 8.0 \um\ we used [5.8] instead of [8.0] and placed an arrow.
The inset shows that all sources in the box 
are distributed along the IR loop,
although the loop of YSOs is quite distorted.
Their colors 
are different from those of previously studied YSOs in nearby star-forming regions. 
We show this by drawing boxes (dotted lines) in Fig. 2
representing the areas occupied by 
young embedded protostars with both circumstellar 
disks and envelopes (Class 0/I) and   
pre-main sequence stars with significant circumstellar disks (Class II)
determined by previous studies \citep[e.g.,][]{muz04}. 
It is clear that the YSOs in the IR loop do 
not fall into the usual Class 0/I or Class II areas and that 
this cannot be due to extinction.
There are many sources 
that fall into and around Class II areas. 
They are outside the IR loop and will not 
be discussed further in this paper. 

Fig. 3 (bottom) shows
a \jmh\ versus \hmk\ color-color diagram for all 110
MIR sources identified in the 2MASS \jhk\ bands. 
The small dots represent sources not detected in 
\spitzer\ MIPS 24~\um\ or IRAC bands while 
the other symbols have the same meaning as in Fig. 2. 
There are only nine 
filled/half-filled
circles because two YSOs with 
large [3.6]$-$[5.8] (Fig. 2) are not identified in the 2MASS bands.
The dashed line represents the main-sequence (MS),
which is from \cite{bes88} for stars later than A0 and is
based on calculations of the \jhk\ colors from 
Kurucz model atmospheres \citep{cas03} for stars earlier than A0.
The dotted line represents interstellar reddening line of 
an A0 star following the reddening law of \cite{bes88}. 
Note that the YSOs are clustered around the positions of 
reddened OB stars. Another way to see this is from the 
color-magnitude ($K_s$ vs. \hmk) diagram (Fig. 3 top).
In this diagram, the dashed line represents the 
MS at 8 kpc calculated from the Kurucz model atmospheres
and using the relation between effective temperature and radius of MS stars.
We adopt the relation of \cite{mar05} for O-type stars and that of 
\cite{sch82} for stars later than B0. 
For B0 stars, we interpolate the 
two with an adopted temperature of $\teff=30,000$ K. 
The dotted lines represent the reddening lines that bound the YSOs. 
The YSOs are again clustered around the positions of 
reddened early-type stars. 
The YSOs are not aligned along a single MS line, but that may be 
due to slightly different amounts of extinction 
toward individual sources. 
The extinction toward individual objects can be derived if 
we assume that they are on the MS, and it 
ranges from $A_V=6.9$ to 9.2 with  
a mean of $A_V=8.0\pm 0.7$ mag. 
The mean extinction agrees very well with the extinction implied from the X-ray
absorbing H columns to the SNR, i.e., $A_V=8.4\pm 0.5$~mag 
from $N(H)=1.6(\pm 0.1)\times 10^{22}$ cm$^{-2}$ \citep{lu02}.
We conclude that the \jhk\ color-color and color-magnitude 
diagrams are consistent with the YSOs 
being massive pre-main-sequence stars at 8 kpc.
The spectral types determined from their $K_s$ magnitudes in Fig. 3 (top) are
B1.5 to O8, which have $\teff=$23,000 -- 35,000 K, masses of 12--21 \msol, and 
luminosities of (1--10)$\times 10^4$ \lsol.

Fig. 4 shows the SED of the most luminous source in the $K_s$ band
(the brightest one in the southern part of the loop in Fig. 1)
as 
an example of the YSO SEDs.
The filled symbols represent dereddened fluxes using the value 
$A_V=6.9$ mag derived from Fig. 3 
and an interstellar reddening law with $R_V=3.1$ \citep{dra03}.
The dereddened SED is characterized by a star-like SED  
with a MIR/FIR excess at $\simgt 8$~\um, as we could have inferred from Figs. 
 2 and 3.
The spectrum at $\simlt 5$~\um\ matches reasonably well 
the SED of  a
35,000 K MS star, the temperature of which was determined from its $K_s$ 
magnitude (Fig. 3). 
This spectrum is quite similar to the SED of early 
Herbig Be (HBe) stars (spectral type B0-B5; group III) \citep{hil92, mal98}. 
A dip at 6--10 $\mu$m similar to that in Fig. 4 was observed in HBe stars and 
has been interpreted as a physical hole 
in the dust distribution caused by the break-up of inner disks
\citep{mal98}. 
If we consider the MIR/FIR excess at $\simgt 8$~\um\ 
to arise from an optically thick disk with a constant effective temperature and 
fit the SED with a 35,000 K star plus a blackbody, we 
obtain a disk with a temperature of 190 K and a luminosity of 
$160$ \lsol, although this fit is not quite satisfactory.
Most YSOs in the IR loop 
have SEDs similar to Fig. 4, 
with some objects showing weak 
MIR excess also. 
The ratio of the disk to stellar luminosity is $\simlt$ 1\%,
showing that there is not much circumstellar material near the YSOs.
If we use the empirical relationship of \cite{nak05} 
between stellar masses and $K$ magnitudes for group III HBe stars, 
the masses of the YSOs range from 12 to 18~\msol\ 
which agrees with our estimates above.
The 2MASS survey has a limiting $K_s$-band sensitivity of 14.3 mag,
corresponding to a 6 \msol\ star at 8 kpc with 8 mag of extinction; no
additional MIR sources are seen between 13.2 and 14.3 mag 
in the vicinity of the IR loop, however,
which suggests a non-standard IMF for this region.

\section{SNR/IR Loop Association and Their Distance}

The IR loop
surrounds the X-ray/radio bright central portion of the SNR almost 
perfectly. 
The SNR is somewhat extended along the northeast-southwest direction, 
suggesting that its expansion has been blocked toward the IR ring. 
We note that the radio contours are deformed around 
the northwest, which is consistent with encountering a dense medium there. 
The positional coincidence with a detailed spatial correlation  
strongly suggests that the IR loop and the SNR \g54 are associated.
The agreement between the extinction toward the IR loop and the 
X-ray absorbing column toward the SNR 
further supports the association (\S~3).
The absence of direct evidence 
for the interaction of the SNR shock with the dense gas may be understood if 
we assume that 
the IR loop is a partial shell in a low density medium, so that
the SNR shock has been able to propagate 
well beyond the loop now.

The distance to \g54\ has been determined by \cite{cam02}, who
derived the dispersion measure of the central
pulsar (308 cm$^{-3}$ pc) and noted that it corresponded to
a distance of 8--12 kpc, depending on the
electron-distribution model of the Galaxy.
We have obtained an \schi absorption spectrum of the SNR using the VLA
Galactic Plane Survey data \citep{sti06}. 
The \schi spectrum of the SNR shows
\schi absorption over all positive velocities, which indicates
that \g54\ is beyond the tangent point ($\ge 5$~kpc).
The absorption peak occurs at +23 \kms, and 
at this velocity there is a faint $^{13}$CO $J=1$--0 emission 
coincident with the IR loop in 
the Boston-University-Five College Radio Astronomy Observatory Galactic Ring Survey 
\citep{jac06}.
The emission is weak ($T_b\le 0.7$ K) 
and implies a mass $\simlt 400$~\msol.
The kinematic distance corresponding to +23 \kms\ is
8.2 kpc using the rotation curve of \cite{bra93}, who
adopt $R_0=8.5$~kpc and $\Theta_0=220$~\kms.
This is consistent with the distance derived from the pulsar DM.
We adopt 8 kpc as the distance to \g54\ and the IR loop, and 
note that this is the distance
to the Perseus spiral arm in this direction. 

\section{Post Main-Sequence Triggered Star Formation?}

The alignment of the
YSOs along a small, loop-like structure strongly suggests 
that we are observing star formation triggered by some mechanism
originating at the center. It cannot be the 
SNR \g54, however, because the remnant is only a few thousand years old.
Instead, the location of the \g54\ pulsar near the center of the IR loop 
indicates that it is most likely the progenitor star of the SNR and that 
the massive progenitor star produced an HII region/wind bubble that
compressed the ambient medium into a shell that fragmented and collapsed
to form the YSOs. 

What is more
interesting about this system is that the triggered
star formation must have occurred near the end of the progenitor's life,
possibly
when it was in its post main-sequence phase.
The ages of early Herbig Be stars are not well known because they usually cannot
be estimated from the H-R diagram \citep[e.g.,][]{tes98}. 
The surrounding diffuse emission in the IR loop, which might be the emission
from dust grains in natal cloud heated either by the YSOs or/and other
nearby B stars, indicates that they are young. 
We can set a limit on the age of the YSOs by noting that 100 \msol\ of stars
in a region with a radius of 2 pc should have a 
1D velocity dispersion of
about 0.2 ~\kms. In order for these stars to 
have a regular spacing of about 0.7 pc
as observed, they could not have moved more than $\sim 0.35$~pc, which sets  a limit
on their age of about 2 Myr.
This limit is significantly shorter than the main-sequence lifetime of 
the progenitor star:
\citet{che05} identified \g54 as the remnant of Type IIP SN, with a progenitor 
mass $\sim 10-25 M_\odot$. 
The destructive impact of a massive star on the ambient molecular
cloud increases rapidly with stellar mass \citep{che99}, so the
small size of the IR loop and the regular
structure seen in the spatial distribution of the YSOs suggest
that the progenitor was near the lower end of this mass range.
A $15 M_\odot$ star has a lifetime $\sim 13$~Myr, and it 
spends about 10\% of its life in the post main-sequence phase \citep{sch92}.
The post main-sequence lifetime increases for lower stellar masses. 
Therefore, 
it is possible that the 
YSOs formed
during the post main-sequence phase,
although it is likely that the progenitor had a significant
dynamical effect on the
natal cloud while it was still on the main sequence.

	We do not have enough information to determine why the triggering
was delayed, i.e., why it did not occur during the first few Myr as 
theories predict \citep[e.g.,][]{hos06}.
Two possibilities can be ruled out:
The triggering was most likely not due to the
wind associated with the red supergiant (RSG) phase of evolution, since 
a typical RSG wind
is far too weak to have a significant effect on the shell \citep{che89}.
In any case, the RSG wind in \g54 may have been particularly weak since no
SNR shock propagating into ambient or circumstellar medium has 
been detected \citep{che05}.
One might also think that the temperature and ionization of the ambient
molecular gas would drop significantly when the progenitor evolved off
the main sequence, but 
\citet{spa94} showed that the temperature and ionization in molecular
gas that is near a star do not vary much for 
effective temperatures in the range $(6-30)\times 10^3$~K. 
A potential explanation for the delay 
in the triggering of the star formation
is that the natal
cloud was shielded by other material until a few Myr ago, when
it was exposed to the star by a combination of motion of
the star relative to the cloud and photoevaporation of the
intervening gas.  Future 
observations will reveal the properties of the newly formed stars and
shed light on the triggering of star formation by evolved stars.

\acknowledgements 
This work is based in part 
on observations with \akari, a JAXA project with the participation of ESA, and also 
in part on observations made with the {\em Spitzer Space
Telescope}, which is operated by the Jet Propulsion Laboratory, California
Institute of Technology under a contract with NASA.
We wish to thank David Hollenbach and Stan Kurtz for their
helpful comments.
This work was supported by the Korea Science and Engineering Foundation 
(R01-2007-000-20336-0) and the Korea Research Foundation (R14-2002-058-01003-0).
The research of CFM is supported by the National Science Foundation through
grants AST-0606831 and PHY05-51164.
This publication makes use of molecular line data from the Boston 
University-FCRAO Galactic Ring Survey (GRS). 

{}

\clearpage

\begin{figure}
\epsscale{1.0}
\plotone{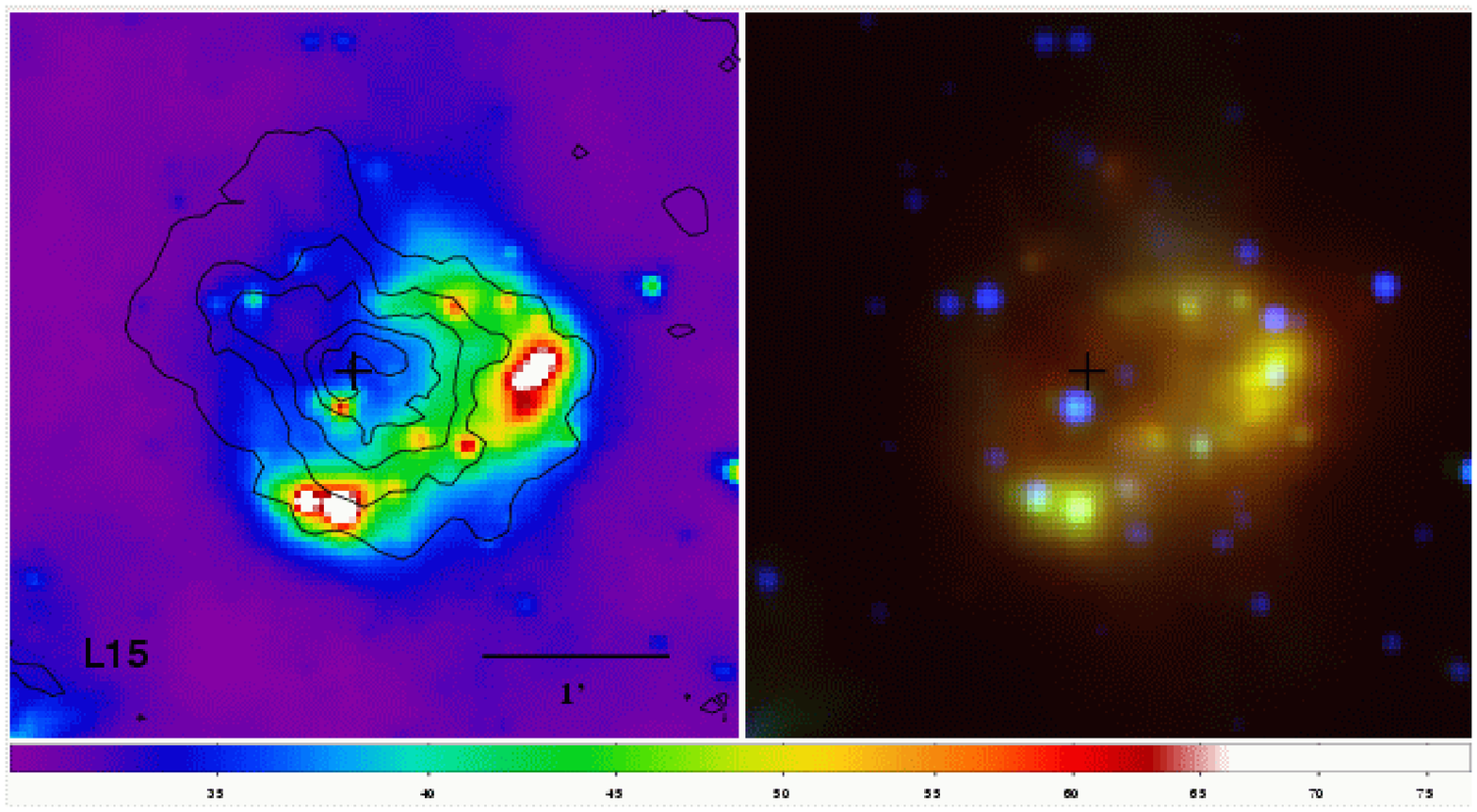}
\caption{
(left) \akari 15 $\mu$m image of the SNR \g54.
The cross marks the position of the pulsar
at $(\alphaJ,\deltaJ)=(19^{\rm h} 30^{\rm m} 30^{\rm s}.13, +18^\circ 52' 14.''1)$.
VLA 4.85 GHz radio contours are overlaid.
The contour levels are equally spaced at every 1.28 K from 0.22 K in brightness temperature. 
Contour levels increase toward the center of the SNR.  
The scale bar corresponds to $1'$ or 2.3 pc at 8 kpc. 
The color bar is given at the bottom of the image in units of 
MJy sr$^{-1}$. North is up and east is to the left.
(right) Three-color image generated from 
the \spitzer\ IRAC 5.8 \um\ (B), the \akari 15 \um\ (G), and 
the \spitzer\ MIPS 24 \um\ (R). 
}

\epsscale{0.8}
\plotone{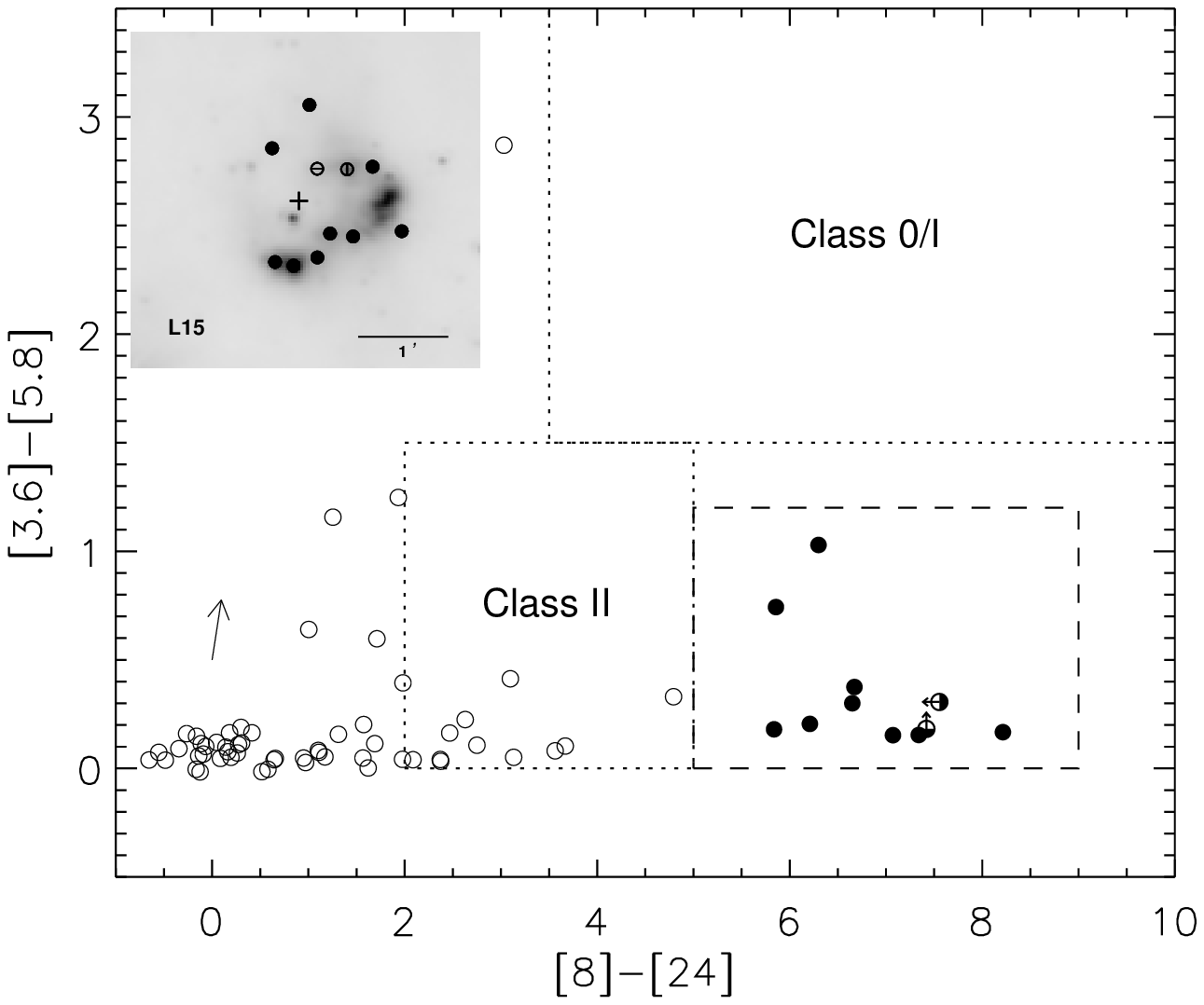}
\caption{\spitzer\ color-color diagram for 64 MIR sources  
detected in the IRAC and MIPS 24 \um\ bands
with a [24] magnitude error less than 0.2 mag.
The sources along the IR loop are well defined in this diagram
by their strong \emt\ excess and moderate or small \tmf\ excess, i.e.,
$5<$\emt$<9$ and $0<$\tmf$<1.2$. The sources within this box are marked by 
solid dots. (See the text for an explanation of half-filled dots.)
The areas occupied by class 0/I and class II sources 
determined from previous studies are shown as dotted lines. 
The arrow is a reddening vector of $A_V=10$.
The inset shows the distribution of the sources within the dashed 
box on the \akari\ 15 \um\ image.
}
\end{figure}

\begin{figure}
\epsscale{0.8}
\plotone{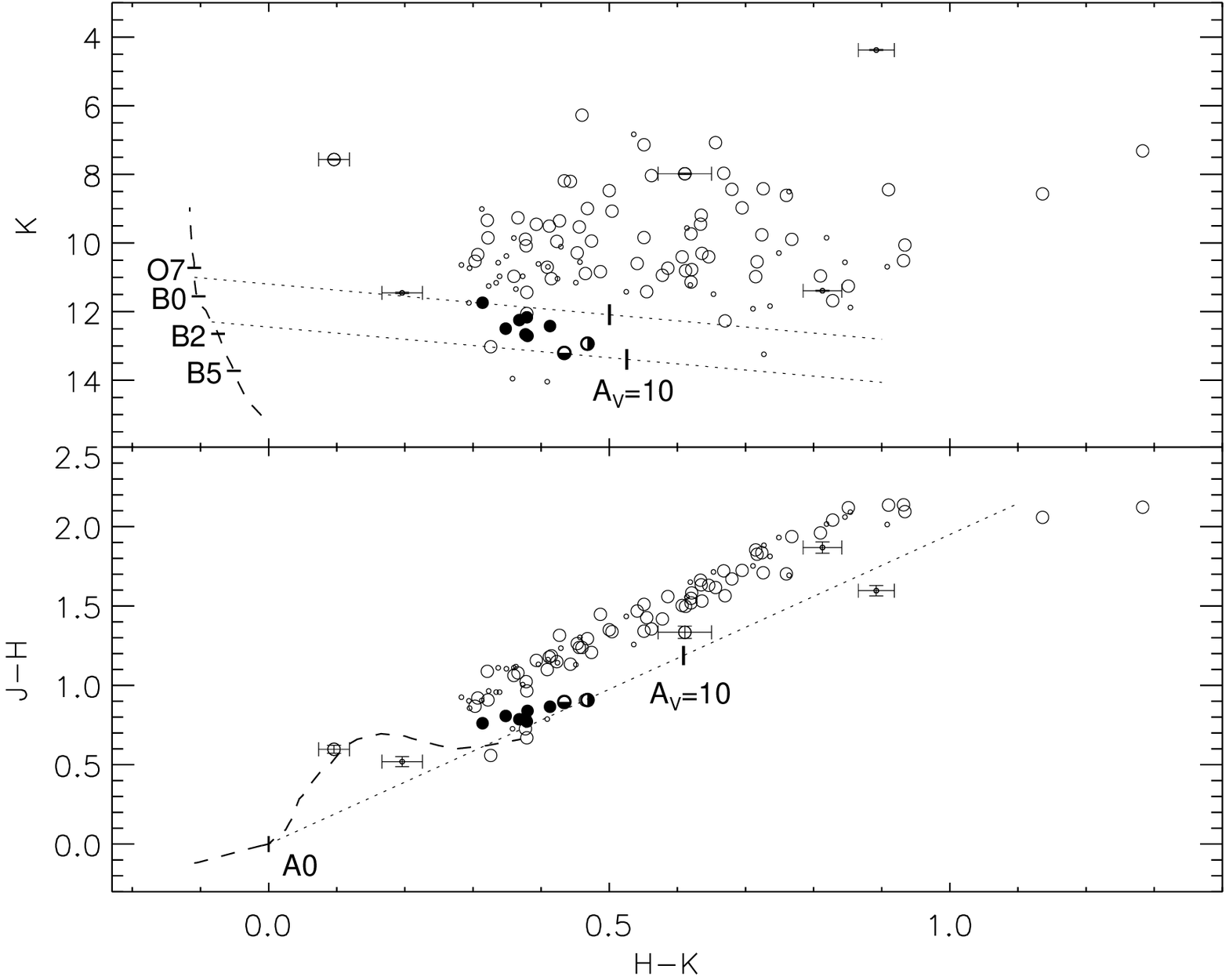}
\caption{ 
The \tmass\ \jhk\ color-color and color-magnitude diagrams 
for all 110 MIR sources identified in the 2MASS \jhk\ bands.
The small dots represent sources not detected in either
\spitzer\ MIPS or IRAC bands while
the other symbols have the same meaning as in Fig. 2. The dashed line 
represents the MS. 
In the color-color diagram, the dotted line represents interstellar reddening line of 
A0 star. Along the reddening line, the positon at which $A_V=10$ is marked.
In the color-magnitude diagram, the dashed line with spectral types 
represents the MS at a distance of 8 kpc.
The dotted lines represent the reddending lines that bound the YSOs. 
Note that the YSOs (solid dots) are clustered around the positions of 
reddened B1.5--O8 stars. 
}

\epsscale{0.8}
\plotone{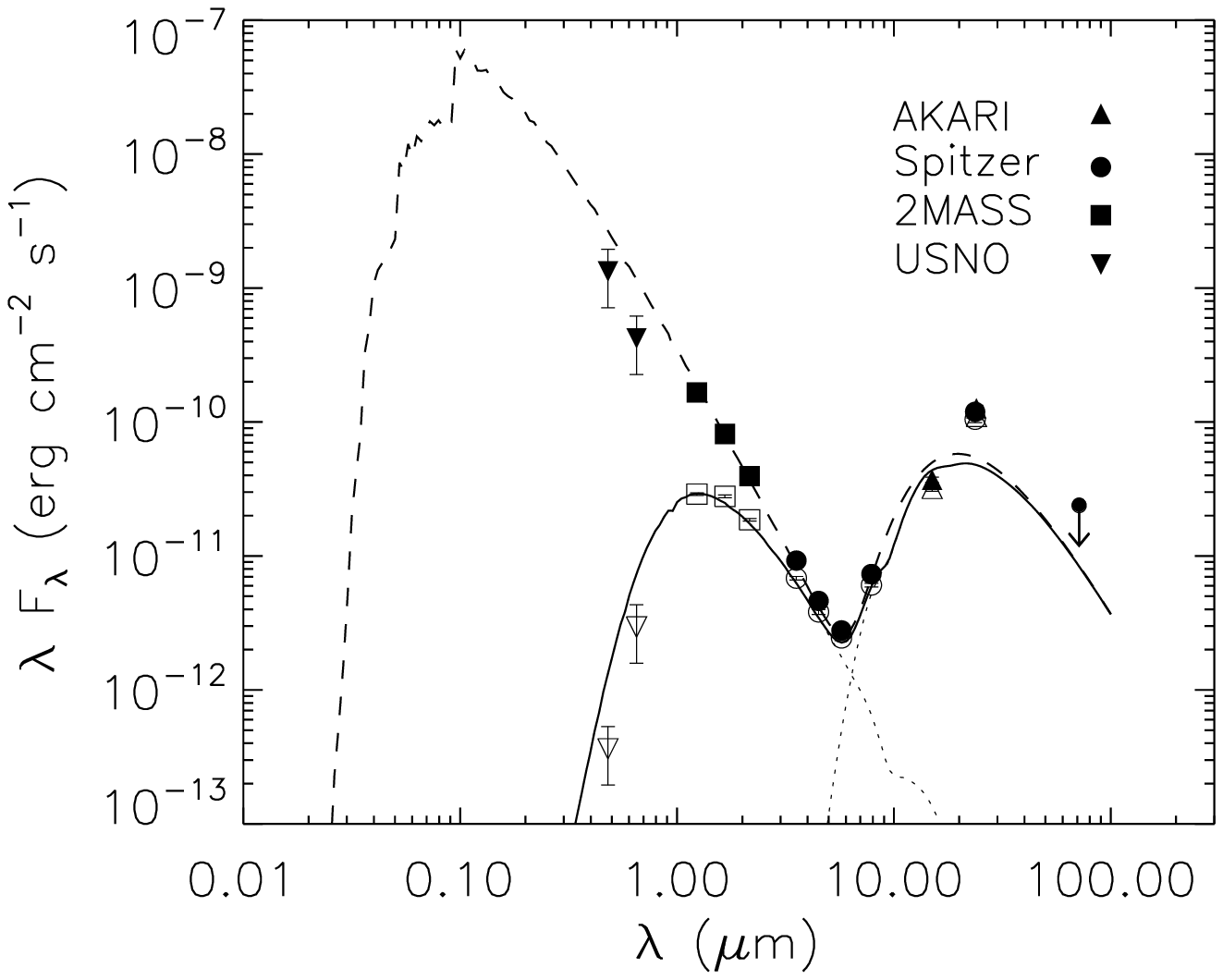}
\caption{ SED of the most luminous YSO in the $K_s$ band. The open 
symbols represent 
observed fluxes while the filled symbols represent extinction-corrected 
fluxes. The \spitzer\ 70 $\mu$m flux is an upper limit. 
The dashed line is a model SED composed of a stellar emission 
at $T_*=35,000$ K and a black body at 190 K. 
The solid line shows the reddened SED and the dotted 
lines their individual components.
}

\end{figure}

\end{document}